# Coherent Power Scaling in Photonic Crystal Surface Emitting Laser Arrays

Ben C. King[1], Katherine J. Rae[1], Adam F. McKenzie[1,2], Aleksandr Boldin[1], Daehyun Kim[1], Neil D. Gerrard[1], Guangrui Li[1],

Kenichi Nishi[3], Keizo Takemasa[3], Mitsuru Sugawara[3], Richard J.E. Taylor[4], David T.D. Childs[1], Richard A. Hogg[1]

[1]James Watt School of Engineering, University of Glasgow, UK

[2]CST Global Ltd., UK

[3]QD Laser Inc., Japan

[4]Vector Photonics Ltd., UK

**Abstract**

A key benefit of photonic crystal surface emitting lasers (PCSELs) is the abillity to increase output power through scaling the emission area while mainting high quality single mode emission, allowing them to close the brightness gap which exists between semiconductor lasers and gas and fibre lasers. However, there are practical limits to the size, and hence power, of an individual PCSEL device and there are trade-offs between single-mode stability and parasitic in-plane losses with increasing device size.  In this paper we discuss 2D coherent arrays as an approach to area and coherent power scaling of PCSELs. We demonstrate in two and three element PCSEL arrays an increase in the differential efficiency of the system due to a reduction in in-plane loss.

Photonic crystal surface emitting lasers (PCSELs) are a new class of laser diode which incorporate a 2D photonic crystal (PC) layer into a semiconductor laser structure. They offer high power single mode surface emission with narrow divergence[1], and control over wavelength, polarization[2], emission beam shape[3], and on-chip beam-steering[4]. PCSELs allow large single-mode powers through scaling the emission area[5].

For a PCSEL, the group velocity of light becomes zero at the band-edge which results in the formation of large and stable two-dimensional single-cavity modes. At these lasing points, waves propagating in certain directions couple, increasing the mode density. Wave coupling is possible according to the Bragg condition, resulting in vertical emission, orthogonal coupling, and 1D scattering (reflection). Figure 1 a) is a schematic of a PCSEL highlighting the key optical loss mechanisms; internal loss $\alpha_i$, $\alpha_\perp$ the out-of-plane emission analogous to the mirror loss in a Fabry-Perot laser, and $\alpha_{//}$, parasitic loss associated with in-plane loss of optical power[5]. The value of $\alpha_{//}$ increases exponentially with reducing PC atom number, and may be so significant as to make lasing impossible to achieve in small devices[7,8] unless the light is confined in-plane[9], See supporting information S1. It is clear that for the ideal PCSEL $\alpha_i$ and $\alpha_{//}$ are minimized and $\alpha_\perp$ is optimized, e.g. high differential output powers and modest threshold gain values to reduce thermal issues within the laser. A simple strategy is to make the PCSEL as large as possible to reduce $\alpha_{//}$ but there is a practical limit to such scaling. These limits include; non-uniform temperature profiles across the device and high cavity temperatures due to self-heating, difficulties in achieving uniform carrier distributions, optical loss and thermal lensing in current spreading layers, and limitations of the scale of e-beam lithography write-fields before stitching errors will be introduced. Current strategies to increase single mode power are to "flatten the mode" by PC design to reduce 1D scattering[6]. This in turn results in increased loss of power in-plane that is currently ameliorated by making the PCSEL of very large area. Here we describe coherent PCSEL array power scaling (See Figure 1) where higher differential efficiencies are demonstrated as the array number increases due to reduced $\alpha_{//}$. In this approach individual PCSELs may be optimized (for e.g. single-mode stability, power per unit area, etc.) with area scaling (and brightness enhancement) achieved through coherent coupling of the array, *reducing* the parasitic losses.

A schematic of the PCSEL devices considered in this paper is shown in Figure 1 b). A PC slab is etched into the GaAs region of a GaAs/InGaAs MQW base wafer and subsequently re-grown so as to create a void within the semiconductor matrix, as shown in the cross-sectional TEM of Figure 1 b). Details of the fabrication process are described in the methods section. The voids provide a high coupling strength in the PCSEL due to the large index contrast between air and semiconductor, albeit with reduced mode overlap with the PC as compared to all-semiconductor counterparts[10,11]. Each PCSEL element is 150um x 150um (480x480 periods), and each PCSEL element is connected by a 150um x 1000um contacted coupler region. The coupler regions are the same structure as a PCSEL element, but without the PC definition process. This allows for these regions to be electrically driven from loss, through transparency, and into gain in order to control the interaction of adjacent elements. This allows individual PCSEL elements to interact through the light emitted at the edges of the devices.

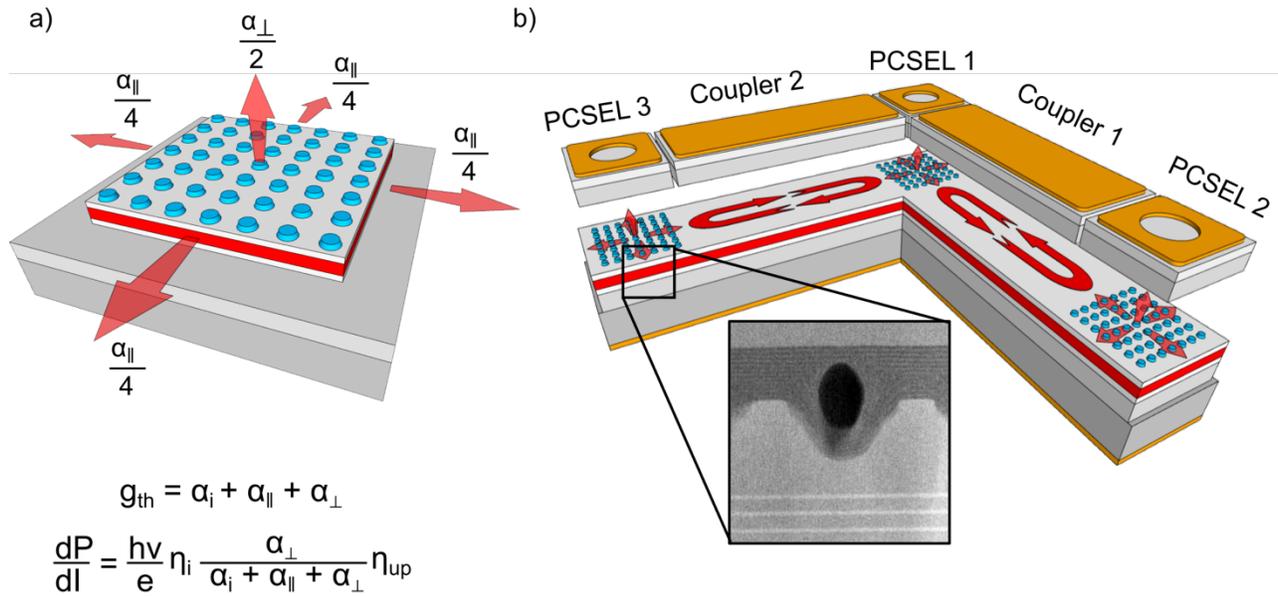

Figure 1 a) Schematic of loss mechanisms in PCSELs b) Schematic of the PCSEL array considered in this paper, where three PCSEL devices are connected by contacted coupling waveguides. Inset: TEM image of PC region of a regrown PCSEL.

Figure 2 is a plot of the LI characteristics of each of the individual PCSEL elements, demonstrating a lasing threshold of ~60 mA, and individual powers of several mW at 300 mA. In our PCSEL design half of the PCSEL power is lost to the substrate, and a third of vertical power is absorbed by the p-side metal contact, these issues can be improved through using substrate emission and constructively utilizing reflection from the top-contact[5]. The lasing wavelengths of the three individual elements are 1067.85±0.25nm. The inset to Figure 2 is a plot of the measured sub threshold photonic band-structure of an individual PCSEL device, overlaid with a simulated band-structure focused around the 2nd order Γ point using structural information from TEM in Figure 1. An excellent agreement between simulation and experiment is obtained. See Methods section.

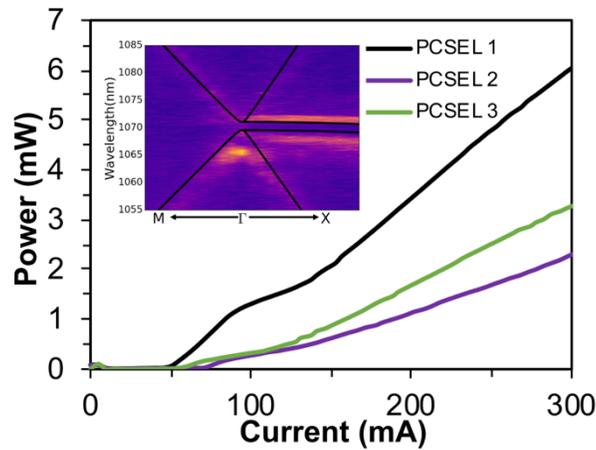

Figure 2: LI characteristics of the individual PCSEL elements. Inset: Measured photonic bandstructure of our device in Γ-X and Γ-M directions overlaid with simulated results (black), focused around the 2nd order Γ point.

Initially we consider power scaling in a two element PCSEL array. Figure 3 considers two PCSEL devices, Element 1 and Element 2, and their connecting coupler region, Coupler 1 The solid lines are plots of the total power of light emitted from the top surface of the PCSELs, as the current to Coupler 1 is varied. The symbols plot the differential efficiency of the system. It can be seen that when there is no current supplied to the PCSELs (blue), where the PCSELs are acting as lossy mirrors, there is little emission from the system until the system begins to lase at ~250mA with the same emission wavelength for each of the two PCSEL elements. When the PCSELs are held at 300 mA (red) the initial power is the sum of the power from each PCSEL at 300 mA (~250 mA above threshold). Both elements are lasing at slightly different wavelengths, separated by 0.3nm. See S3. As the coupler current is increased, there appears to be a soft turn-on at ~150mA, seen more clearly in the plot of differential efficiency. Further characterization of the device indicates a transparency current for the coupler of 239mA. See S4. At this coupler current, the differential efficiency is 0.026 W/A, higher than that of the average of the two PCSELs (c.f. 0.020 W/A).

The total power of the system at coupler transparency is 10.75 mW, an enhancement of 26% compared to the initial PCSEL power with no coupler current. The differential efficiency of the system driven in this way rises to a maximum of 0.037 W/A at 400mA. When the coupler is above transparency, the emission from the two PCSEL elements is observed to be identical. See S3. This coincidence in wavelength is characteristic of phase-locking and coherence of the PCSEL elements[12,13].

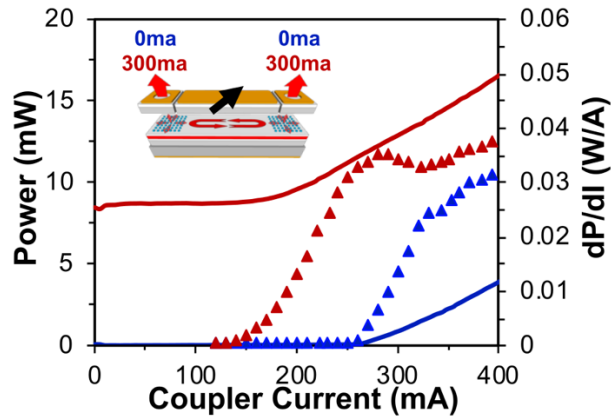

Figure 3: Line plots show the power of light emitted from PCSEL Elements 1 and 2 for varying Coupler 1 current when there is no current supplied to PCSEL devices (blue) and when 300ma is supplied to each device (red). Scatter plot shows the differential efficiency of the system.

Figure 4 describes similar results for a system of three PCSEL elements connected as shown in the inset, with an addition PCSEL Element 3connected to PCSEL Element 1 through Coupler 2. The solid lines are again plots of the total power of light emitted from the surface of the three PCSEL elements, as the currents to the Coupler 1 and Coupler 2 are concurrently varied. The symbols again plot the differential efficiency of the system. Lasing of the system with zero PCSEL current is once more observed at ~250mA. When all three PCSELs are held at 300 mA (red) the initial power is the sum of the power from each of the three lasing PCSELs. As compared to the two PCSEL element system, an even softer "turn-on" is observed with a differential efficiency of 0.026 W/A at the transparency point for the couplers (c.f. 0.019 as the average for three independent PCSEL elements and 0.020 for two coupled PCSEL elements). A maximal differential efficiency of 0.028 W/A is observed at ~350 mA. The total power of the system at coupler transparency is 15.9 mW, an enhancement of 34% as compared to the three individual PCSELs alone.

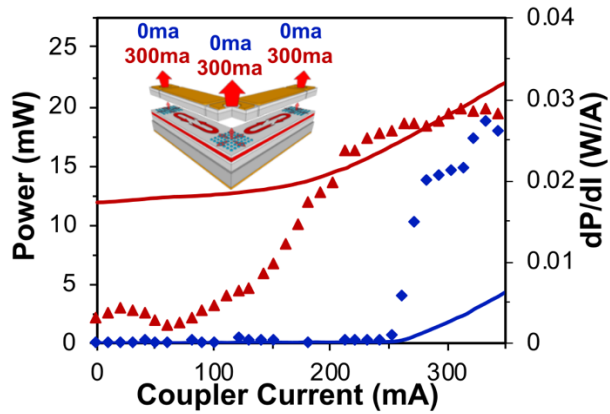

Figure 4: Line plot shows the power of light emitted from PCSEL 1, 2, and 3 for varying Coupler 1 and 2 current when there is with no current being supplied to the PCSEL devices (blue) and when 300ma supplied to each device (red). Scatter plot shows the differential efficiency of the system.

The soft turn-on observed in the two and three element PCSEL arrays is attributed to the gradual increase in transparency of the coupler regions with increasing current and appears softer in the three element array due to higher power within the system. The increase in differential efficiency at transparency for the couplers is of particular interest in determining the change in $\alpha_{//}$. For these PCSEL elements $\alpha_{//}$ is calculated to be 24cm$^{-1}$ (See S1), and $\alpha_i = 3$cm$^{-1}$. With an internal quantum efficiency of 0.9, we obtain a value of $\alpha_\perp$ of 1.5 cm$^{-1}$ to achieve 0.02 W/A. Utilising the ratios of dP/dI we can deduce a new effective $\alpha_{//}$ for the two element array of 17.4 ±0.5 cm$^{-1}$ and the three element array of 16.3 ±0.5 cm$^{-1}$. This is in agreement with predictions that we should achieve 18 cm$^{-1}$ and 16 cm$^{-1}$, respectively. See S5. We note that scaling the array leads to the effective $\alpha_{//}$ being proportional to $1/n$ (where $n$ is the PCSEL array order). As the coupler current is increased past transparency, not only is there a benefit due to the re-cycling of optical power lost in-plane, but also amplification of this light within the coupler that provides additional power.

It has been pointed out that if power per unit area and single-mode operation can be maintained, then scaling PCSEL area results in a proportional increase in brightness[6]. PC design to enhance $\alpha_\perp$ along with area scaling of device size resulted in Watt-level single-mode emission[5], but single-mode power (and brightness) was limited by the onset of multi-modal lasing, with this being area limited. Subsequently, record single-mode powers being achieved utilizing a PC design through increased threshold gain difference between fundamental and higher order modes[6], allowing increased single mode area scaling. In these PC designs, as gain difference is increased, so $\alpha_{//}$ increases and $\alpha_\perp$ decreases. A key trade-off is therefore observed between single-mode stability and parasitic losses. By contrast, in our approach area scaling is achieved through coherent coupling of the array *reducing* the parasitic $\alpha_{//}$ losses.

**Methods**

The MQW sample (base wafer) was grown via molecular beam epitaxy on a 3" (100) n-GaAs substrate, grown 2° off towards $1\bar{1}0$. It consists of, from bottom to top, a 1400 nm n-$Al_{0.4}Ga_{0.6}As$ cladding layer, 30 nm GaAs, a 3x GaAs/InGaAs quantum well active region consisting of 6.5 nm InGaAs and 31.5 nm GaAs barriers, and finally 160 nm of p-GaAs. A 2D PC slab is etched into this top GaAs layer through RIE. S2 shows a cross sectional SEM of a similar PCSEL device after PC definition. The wafer is then cleaned through UV ozone treatment and an HF clean before regrowth in an MOVPE reactor. The regrowth consisted of 125 nm of AlAs, alternating with 1 nm of GaAs ever 9 nm of AlAs, 1500 nm p-$Al_{0.4}Ga_{0.6}As$ cladding layer, and highly doped p++ GaAs cap layer. Following regrowth PCSEL and coupler regions were isolated through etching of the top p++ cap layer to form isolated mesas . A 200 nm-thick $SiO_2$ passivation layer was then deposited across the wafer, into which seperate contact windows were opened for the individual PCSELs and couplers, and *p*-type Ti/Pt/Au contacts deposited. For the PCSELs a 60 nm-diameter circular emission aperture was defined in the top contact. A backside Ni/Au/Ge/Ni/Au n-type contact was then deposited to form the finished devices.

The bandstructure of our device was mapped by scanning a 200 um multimode fibre across collimated sub-theshold emission from from the surface of the PCSEL, scanning incrementally in the $\Gamma$-X and $\Gamma$-M directions to build up a bandstructure around the 2$^{nd}$ order $\Gamma$ point[14]. Overlaid is a simulate photonic bandstructure, simulated through plane wave expansion methods.

**Acknowledgements**

This work was supported by the Engineering and Physical Sciences Research Council (RCUK Grant No. EP/L015323/1).

**Data Availability Statement**

The data that support the findings of this study are available from the corresponding author upon reasonable request.